# Domain wall propagation in Permalloy nanowires with a thickness gradient

O. Petracic[a)], P. Szary and H. Zabel

[a]Institut für Experimentalphysik, Ruhr-Universität Bochum, D-44780 Bochum, Germany

D. Görlitz and K. Nielsch

[b]Institute of Applied Physics and Microstructure Research Center, University of Hamburg, D-20355 Hamburg, Germany

The domain wall nucleation and motion processes in Permalloy nanowires with a thickness gradient along the nanowire axis have been studied. Nanowires with widths, $w$ = 250 nm to 3 µm and a base thickness of $t$ = 10 nm were fabricated by electron-beam lithography. The magnetization hysteresis loops measured on individual nanowires are compared to corresponding nanowires without a thickness gradient. The $H_c$ vs. $t/w$ curves of wires with and without a thickness gradient are discussed and compared to micromagnetic simulations. We find a metastability regime at values of $w$, where a transformation from transverse to vortex domain wall type is expected.

The propagation of magnetic domain walls (DWs) in confined systems is of strong interest due to its relevance in data storage and processing applications [1, 2] and due to novel behaviors exclusively found in nanoscale objects. E.g. only in nanowires one encounters head-to-head transverse (TDWs) or vortex domain walls (VDWs). The occurrence of which depends on the nanowire width, $w$, and thickness, $t$ [3-12]. A $w$-$t$-phase diagram for Permalloy ($Ni_{81}Fe_{19}$) nanowires has been explored both experimentally [9] and theoretically [5, 11] showing the regions of existence of either TDWs or VDWs. Generally, narrow and thin wires exhibit TDWs, whereas wider and thicker wires show VDWs. Transformations between these two configurations were observed either upon heating [9] or by application of an electrical

---

[a)] Corresponding author: oleg.petracic@ruhr-uni-bochum.de

current [10]. However, DW transformations triggered by the wire geometry are not well explored.

In order to study the latter case, straight Permalloy (Py) nanowires with a thickness gradient along the wire axis have been prepared. We use electron beam lithography and subsequent lift-off for fabrication. The metallization has been performed by UHV ion-beam sputtering from a Py target. The wires have widths in the range 250 nm $\leq w \leq$ 3 μm, a base thickness of $t$ = 10 nm and a length of 90 μm with laterally tapered ends. The tapered ends have the effect of eliminating unwanted end domains. Sets of identical wires but different lengths of thickness gradients $0 \leq l_g \leq$ 40 μm with a constant slope of $\approx 5 \cdot 10^{-4}$ have been fabricated using a penumbra shadow mask during metallization very close to the substrate [13]. The idea is to trigger a transformation of a DW from TDW to VDW type at the thickness gradient. Depending on whether a transformation occurs or not the remagnetization behavior of the wire will show different signatures.

Figure 1 shows scanning electron microscope (SEM) images of a wire with $w$ = 3 μm without a gradient (a) and a similar wire with a gradient (b). The gradients have an almost linear thickness profile as found by atomic force microscopy (AFM) studies (not shown). Magnetometry measurements have been performed using a 'NanoMoke II' setup (Durham Magneto Optics) [14], where the longitudinal magneto-optic Kerr effect (LMOKE) signal of *individual* nanowires can be recorded as function of the applied field.

Figure 2 (a) shows example LMOKE hysteresis loops for wires with $w$ = 250 nm without gradient (label 1) and with a gradient (label 2). In all cases we find square loops indicating magnetization switching via DW nucleation and motion. Furthermore, one observes a strong decrease of $H_c$ upon introducing a thickness gradient. In detail, one finds for wires without a gradient $H_c$ = 104, 78, 55, 31, 15 and 9 Oe for widths $w$ = 250 nm, 350 nm, 530 nm, 800 nm, 1.5 μm and 3.0 μm, respectively, as shown in Fig. 2 (b) [black circles, together with a linear fit]. The width dependence matches well with the expected linear behavior, $H_c = c_1 + c_2 t/w$, where $c_1 \propto H_c(w \to \infty)$ is a usually negligible constant and $c_2 \propto M_s$ another constant depending also on the specific wire shape [7, 12]. The Laser spot (diameter < 5 μm ) has been focused onto the middle part of the wire. Corresponding wires *with* a gradient of length $l_g \approx$ 40 μm show reduced values, i.e., $H_c$ = 28, 11, 8.4, 7.7, 4.7, and 3.0 Oe, respectively, as shown in Fig. 2 (b) [blue stars, the interpolating line is a guide to the eye]. Here the Laser spot has been focused onto the wire in a distance of ≈10 μm next to the left end of the wire. The two curves $H_c$ vs. $t/w$ seem to converge with two different slopes for $t/w \to 0$ (increasing $w$). At $t/w > 0.01$ the slope of the curve flattens and a plateau or inflection point is visible at $t/w \approx$ 0.017. For greater values, $t/w > 0.02$, the curve shows an increasing slope for $t/w \to \infty$ (decreasing $w$).

Considering the *w-t*-phase diagram for Py nanowires [9] one expects a TDW-VDW transition at $w \approx 450$ nm for a base wire thickness, as in our case, of $t = 10$ nm and hence $t/w \approx$ 0.022. This value matches relatively well with the inflection point of the curve. Obviously it marks a *crossover* between two different regimes of domain wall propagation through the gradient. I.e., in the first regime for $w < 450$ nm ($t/w > 0.022$) it follows from the *w-t*-diagram and the relation $H_c \propto t/w$ that a TDW will nucleate in the thinnest part of the wire. Since it is expected to stay a TDW even in the thick part, no transformation of the type of DW will occur and it will propagate unhindered through the entire wire. This is reflected in the slope of the curve for large $t/w$ which seem to approach the slope of the curve for wires without gradient. In the second regime for $w > 450$ nm ($t/w < 0.022$) again a TDW is expected to nucleate in the thin part. However, during propagation over the thickness gradient it has to transform to a VDW. This transformation most probably occurs over an energy barrier thus inhibiting the DW propagation. This is reflected in the $H_c$ vs. $t/w$ curve as the part with reduced slope for $t/w \to 0$ compared to curve 1 of the wires without gradient. The range between the two regimes, $t/w \sim 0.017$, is most likely governed by a bifurcation between the two cases.

Furthermore, we performed micromagnetic simulations using OOMMF [15]. Fig. 3 shows $H_c$ vs. $t/w$ for wires without a thickness gradient (label 1, squares) and with a gradient (label 2, diamonds). The simulated wires had widths $w$ = 50, 70, 100, 200, 300, 400 and 500 nm and a base thickness of $t$ = 10 nm. The lengths have been set to $l = 10 \cdot w$ and the cell size to $[\Delta x, \Delta y, \Delta z]$ = [10, 10, 2.5] nm. Standard micromagnetic parameters for Py have been used, i.e., $M_s$ = 860 kA/m and $A$ = 13 pJ/m. The gradient has been introduced by vertically tapering off the 'filled' simulation cells over a fixed distance of 300 nm. Wires without a gradient show a curve with linear behavior similar to the experimental case. One should note that the fields are larger compared to the experiment probably due to the simulation being at $T$=0 and due to less nucleation seeds being present compared to real samples. The curve for wires *with* a gradient displays also a cross-over behavior similar to the one in the experiment although the plateau is not as pronounced. One observes also an inflection point at $t/w$ = 0.1 which corresponds to $w$ = 100nm. This matches well with the *theoretically* expected TDW-VDW transition border at $w$ = 150nm for $t$ = 10nm [11]. Consequently the signature of a possible TDW-VDW bifurcation can be observed also in simulations.

Surprisingly, however, the corresponding spin structures during simulation do not show any transition to a VDW configuration. Fig. 4 shows snapshots of the spin structure for a wire with $w$ = 400 nm and $t$ = 10 nm. Starting initially at positive saturation the applied field is gradually decreased. At a negative field of $H$ = -80 Oe one observes first nucleation and then propagation of a TDW, which stops at the first step of the gradient [Fig. 4 (b)]. (The lowest terrace is slightly longer compared to the other ones for more controlled nucleation of the DW.) At a field of $H$ = -120 Oe the TDW depins from the step and becomes stretched and deformed [Fig. 4 (c-d)]. No transition to a VDW can be observed for wires even with $w$ = 700 nm (not shown). Only in simulations of thicker wires with e.g. $t$ = 18 nm vortex formation can be found [Fig. 4 (e)]. Hence, the inflection point in the $H_c$ vs. $t/w$ curves in Fig. 2(b) and 3 does not correspond to a wall transformation.

We rather assume that a *metastability regime* occurs for such values of $w$ and $t$, where a transformation from a TDW to VDW is expected energetically. However, the TDW state stays trapped by an activation barrier in a local energy minimum. This probably results in a reduced mobility of the DW at the gradient appearing in the $H_c$ vs. $t/w$ curve as an inflection point [Fig. 2(b) and 3]. In contrast, the regime for small $w$ (large $t/w$) is characterized by a larger mobility since no transformation is expected. Consequently the $H_c$ vs. $t/w$ curve is mainly governed by the DW nucleation field and therefore by the dimensions and the shape of the wire ends where the DWs nucleate. The same applies to wires without a gradient. The behavior for large $w$ (small $t/w$), however, remains unclear. The slope of the $H_c$ vs. $t/w$ curve is larger than in the plateau region, but less compared to the wires without gradient. This could hint toward an extrapolated regime where DW transformation does occur, but where the mobility of the DW is increased compared to the plateau regime.

In summary, we have investigated the domain wall propagation in Permalloy nanowires with a thickness gradient along the wire axis. The hysteresis loop of individual nanowires with and without a thickness gradient of base thickness $t = 10$ nm, widths 250 nm $\leq w \leq$ 3 μm and a length of 90 μm have been recorded by a MOKE setup. We find square loops indicating magnetization switching via DW nucleation and motion. The coercivity $H_c$ vs. $t/w$ shows for wires without a gradient the expected linear behavior. For wires with a gradient the $H_c$ values are reduced. The $H_c$ vs. $t/w$ curve shows a characteristic inflection point at those values of $w$, where a transformation from TDW to VDW type is expected. Corresponding micromagnetic simulations, however, reveal that no transformation occurs. We assume that the inflection point is rather due to a metastability regime with reduced mobility of the DW at the gradient.

The authors thank the Deutsche Forschungsgemeinschaft through SFB 491 and SFB 668 for financial support.

**Figure captions:**

Fig. 1. Scanning electron microscope (SEM) images of nanowires, (a) without thickness gradient, length $l$ = 90 µm and width $w$ = 3 µm and (b) similar wire but with a gradient at the right hand side. The scale corresponds to 20 µm.

Fig. 2. (a) LMOKE vs. $H$ loops for wires with $w$ = 250 nm without a gradient (label 1, black curve) and with a gradient (label 2, blue curve). (b) Coercive fields $H_c$ vs. $t/w$ of wires without a gradient (label 1, black circles) and of similar wires with a gradient (label 2, blue stars). The black straight line is a linear fit to the data points (circles) and the blue line a guide to the eye. The loops in (a) correspond to the data points at $t/w$ = 0.04 in panel (b).

Fig. 3. $H_c$ vs. $t/w$ for simulated wires without gradient (label 1, squares) and corresponding wires with gradient (label 2, diamonds). Lines are guides to the eye.

Fig. 4. (a) Side view of the wire with $w$ = 400 nm and $t$ = 10 nm. (b-d) Snapshots of the spin structure during simulation at various times starting from *positive* saturation and then applying a field of $H$ = -80 Oe (b) and -120 Oe (c-d). For comparison the spin structure of a wire with $t$ = 18nm (e). The magnetization component along the field $M_x$ = $-M_s$, 0, $+M_s$ is color-coded by 'blue-white-red'.

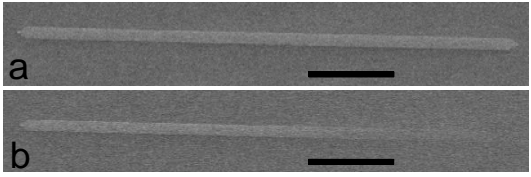

FIG. 1

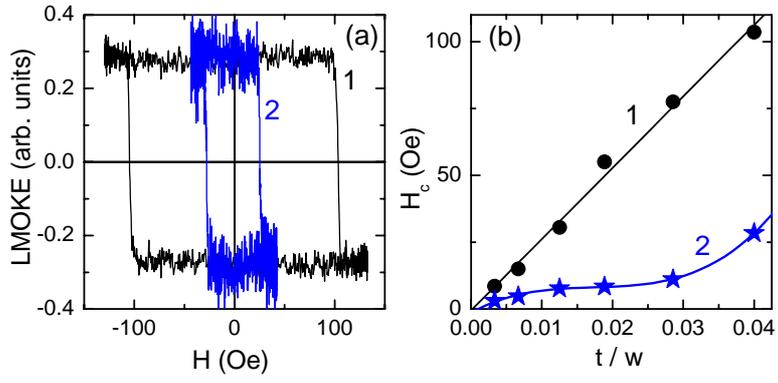

FIG. 2

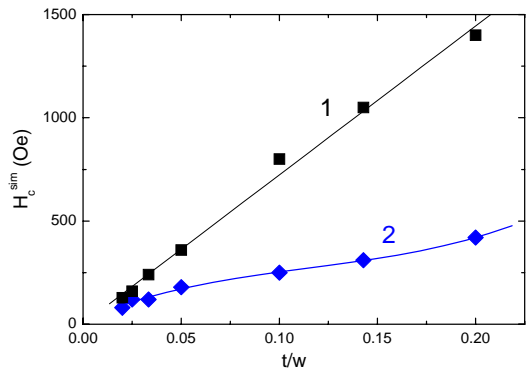

FIG. 3

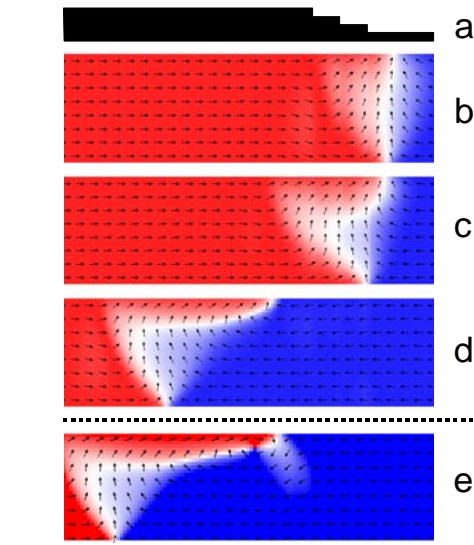

FIG. 4